Quantum Mechanics without Complex Numbers:

A Simple Model for the Electron Wavefunction Including Spin

Alan M. Kadin*

Princeton Junction, NJ

February 22, 2005

Abstract:

A simple real-space model for the electron wavefunction is suggested, based on a transverse wave with helicity, rotating at $\omega = mc^2/\hbar$. The mapping of the real two-dimensional vector phasor to the complex plane permits this to satisfy the standard time-dependent Schrödinger equation. This model is extended to provide an intuitive physical picture of electron spin. Implications of this model are discussed.

*E-mail: amkadin@alumni.princeton.edu

I.        Introduction

Since its inception a century ago, quantum mechanics has generated more than its share of confusion and mystery.  Although its precise applicability was clearly established early on, fundamental issues related to interpretation have continued to be extensively discussed, among both experts and novices.  My motivation here is to focus on one particular aspect, namely the apparently fundamental role of complex numbers in the theory of quantum mechanics.    Specifically, the quite general time-dependent Schrödinger equation takes the standard form

$$i\hbar\ \partial\Psi/\partial t = H\Psi = (-\hbar^2/2m)\nabla^2\Psi + V(r)\ \Psi \qquad\qquad (1)$$

which can only be solved for a complex wavefunction $\Psi = |\Psi|\ \exp(i\phi)$.  As usual $\hbar = h/2\pi$ is Planck's constant, H is the Hamiltonian operator, m is the mass of the "particle", V(r) is its potential energy, $\phi$ is the phase of the wavefunction, and of course $i = \sqrt{(-1)} = \exp(i\pi/2)$.  Furthermore, for an energy eigenstate given by $H\Psi=E\Psi$, the solution takes the form $\Psi(r,t) = \Psi(r)\ \exp(-i\omega t)$, with the rate of phase rotation given by the fundamental Planck relation $E = \hbar\omega$.  The wavefunction intensity $|\Psi|^2$ is understood to represent either a density function or a probability distribution, depending on the interpretation.  The complex phase factor $\exp[i\phi(r,t)]$ provides the basis for quantum interference, but does it describe a physical oscillation or rotation in real space?  The standard interpretation of quantum mechanics does not really address this issue [1].

As quantum mechanics is generally taught, a major focus is on the mathematics of complex Hilbert spaces.  This mathematics is really quite beautiful, but one must not forget that this mathematics is really a representation of the physics, rather than the other way around.  So it is useful to contrast the role of the complex phase in quantum mechanics to that in classical physics.  Classically, the complex phase represents the phase of a real physical sinewave oscillation or rotation.  The complex factor $\exp(i\phi)$ is used just to make the math easier.  In contrast, in quantum mechanics, the full complex wave is generally believed to represent the "real physics", although only real eigenvalues of hermitian operators are "observables".    This would seem to make the complex mathematics an essential part of the physics, rather than just an auxilliary. But is this really necessary?

I would like to suggest, using the electron as an example, that a real two-dimensional transverse wavefunction can provide a simple and appealing physical picture that can account for much of the quantum mechanical phenomenology without the mathematical formalism of Hilbert space. This includes the phenomenon of electron spin, which is typically presented in a rather ad-hoc, even mysterious manner. I am not suggesting that this particular model is the ultimate solution to the interpretation of quantum mechanics, but rather that it provides a straightforward example that may contain some useful insights into the fundamental physics from a somewhat non-standard point of view.

II.     Helical Electron Plane Wave

Consider a free-electron plane wave moving in the z-direction, with energy E and momentum p. Conventionally, this is represented by a scalar complex wave $\Psi_c$ of the form:

$$\Psi_c = \Psi_R + i\Psi_I = \Psi_0 \ \exp[-i(\omega t - kz)], \tag{2}$$

where $\omega$ and $k$ are given by the de Broglie relations $E=\hbar\omega$ and $p=\hbar k$. Since the oscillation is not viewed as a real observable, the zero of energy is arbitrary; any consistent reference energy is acceptable, so that typically $E = \hbar^2 k^2/2m$. There is no obvious way to incorporate electron spin into this picture, except as an additional assertion.

In contrast, I would like to suggest the following alternative representation for the same state, based on a real transverse vector wave $\mathbf{\Psi}$ of the form:

$$\mathbf{\Psi}(r,t) = \Psi_0 \ [\mathbf{u_x} \cos(\omega t - kz) \pm \mathbf{u_y} \sin(\omega t - kz)] = \Psi_x \ \mathbf{u_x} + \Psi_y \ \mathbf{u_y} \ , \tag{3}$$

where $\mathbf{u_x}$ and $\mathbf{u_y}$ are the unit vectors in the x- and y-directions. This is a circularly polarized transverse wave, with either positive or negative helicity depending on whether the plus or minus sign is chosen. For fixed t, the tip of the vector follows a helix (Fig. 1a). For fixed z, one simply has circular rotation at an angular frequency $\omega$ of a vector of length $\Psi_0$ (Fig. 1b). I would suggest that these helical vector waves form a natural representation for electron spin, with the two helicities representing the two spin states of the electron. This is directly analogous to a circularly polarized transverse electromagnetic wave, which carries angular momentum that corresponds in the quantum limit to $\pm\hbar$ per photon.

One can easily transform between these two (complex and vector) representations:

$$\Psi_c = \Psi_x + i\Psi_y; \ \mathbf{\Psi} = \mathbf{u_x}\,\text{Re}(\Psi_c) + \mathbf{u_y}\,\text{Im}(\Psi_c). \tag{4}$$

If one of the two helicities corresponds to a given $\Psi_c$, then the other one corresponds to the complex conjugate $\Psi_c{}^*$. Both conjugate pairs correspond to the same solution to the Schrödinger equation, as would be expected for two spin states. One can also use these mappings to write a 2-component Schrödinger equation in the form:

$$-\hbar\,\partial\Psi_y/\partial t = (-\hbar^2/2m)\nabla^2\Psi_x + V(r)\,\Psi_x\,;$$

$$\hbar\,\partial\Psi_x/\partial t = (-\hbar^2/2m)\nabla^2\Psi_y + V(r)\,\Psi_y. \tag{5}$$

Note that this is now a real two-dimensional vector equation, with no complex numbers.

Thus far the discussion has been limited to a single plane wave, but electrons are generally present in bound states, with standing waves instead of traveling waves. Consider for simplicity the one-dimensional particle-in-a-box, with the electron confined between z=0 and z=L. The solution takes the form of discrete bound states $\mathbf{\Psi_n}$

$$\mathbf{\Psi_n} = \sin(n\pi z/L)(\mathbf{u_x}\cos\omega t \pm \mathbf{u_y}\sin\omega t). \tag{6}$$

Here n=1 corresponds to the ground state and n=2, 3,... to the excited states, and the quantized energies $E_n$ are given as usual by

$$E_n = \hbar\omega_n = \hbar^2 k^2/2m = \hbar^2(n\pi/L)^2/2m \tag{7}$$

and as before the $\pm$ corresponds to the two spin states. Note that this vector wavefunction has separated into two factors, the usual standing-wave envelope and the rotating phase vector. The negative values of the sine (for n>1) correspond to 180º shifts of the rotating phase.

One final adjustment is to the zero of the energy level. If these phase rotations are physically real, then the total energy of the electron cannot have an arbitrary offset. The only way to make it unique, even for a non-relativistic model, is to add the rest mass energy $m_0 c^2$ = 511 keV, corresponding to a frequency $m_0 c^2/h$ ~$10^{20}$ Hz. This is orders of magnitude faster than anything else happening in atomic systems. I will discuss later whether such high-frequency resonant behavior might be directly observable.

### III.    A Hydrodynamic Model for Electron Spin

It is well known that a circularly polarized transverse electromagnetic wave carries angular momentum, corresponding to spin in photons [2, 3]. This should also be true for a circularly polarized transverse electron wave, although the derivation seems less clear for the present case. Ohanian [3] derives spin from physical rotation in the relativistic Dirac equation for the electron. But a simpler (if cruder) picture based on solid-body rotation is used here. For a solid cylinder of uniform mass M and radius R, rotating about the axis of symmetry with angular velocity $\omega$, the moment of inertia is I = ½ $MR^2$, and the angular momentum L is given by

$$L = I\,\omega = \tfrac{1}{2}\,MR^2\,\omega \tag{8}$$

Now consider a hydrodynamic fluid, consisting of a close-packed array of parallel vortices, each of radius R and mass M (see Fig. 2). The total angular momentum density of the fluid is then $L_{tot} = NL_v$, where N is the total number of such vortices. However, note that if the vortex scale R increases by a factor of 2, M increases by a factor of 4, $L_v$ increases by a factor of 16, and $L_{tot}$ increases by a factor of 4. In general, $L_{tot} \propto R^2$. So in order to use this picture for the spin of the electron, we have to identify the scale R of the vortices.

An important consideration is the maximum linear speed of matter on the outside of the vortex, which is $u = R\omega$. Clearly, there is a upper limit to this speed; it can be no larger than c, the speed of light. (In fact, relativistic corrections would be needed for accuracy in this range, but we will be ignoring them here for simplicity.) Taking $\omega = mc^2/\hbar$ and $u_{max} = c$ gives $R_{max} = \hbar/mc \sim 0.4$ pm, the Compton radius $r_c$ of the electron. This is much smaller than typical atomic-scale electron orbitals, so we need to add the contributions of N vortices to get the total spin. For simplicity also assume a 100% packing density of vortices of mass $m_v = m/N$ and ignore intervortex interference. From Eq. (8) above,

$$L_{tot} = NL_v = (N/2)\, m_v\, r_c\, (r_c\omega) = (N/2)\, m_v\, (\hbar/mc)c = (N\hbar/2)\, m_v/m = \hbar/2 \qquad (9)$$

Of course, this has got to be the right answer for the spin of the electron (a spin-½ particle), but it is remarkable that it came out so easily in such a crude model, with no adjustable parameters.

One can also estimate the total energy associated with this spin $E_s$, based on the rigid-body rotational kinetic energy $E_v$ of each vortex:

$$E_s = NE_v = (N/2)I\omega^2 = Nm_v r_c^2 \omega^2/2 = mc^2/2 \qquad (10)$$

While I won't vouch for the exact factor of ½ (again, relativistic and other corrections have been ignored) this suggests that a large fraction of the mass energy of the electron is associated directly with the spin.

One can also estimate the magnetic moment of the electron from this model. Treating the rotating charge per vortex $q_v = e/N$ as a current $i_v = q_v\omega/2\pi$, one obtains

$$\mu = Ni_v A_v = (e\omega/2\pi)(\pi r_c^2) = e\hbar/2m = \mu_B \qquad (11)$$

where $\mu_B$ is the Bohr magneton. Again, this is the correct result, perhaps fortuitously, but it does suggest that this crude model may incorporate much of the essential physics.

IV.     Discussion and Conclusions

To summarize, it has been suggested here that a spatially-rotating vector phasor can provide the physical basis of the quantum mechanical complex phase, as well as producing the spin of the electron.  The crude hydrodynamic vortex model given here for calculating the spin, while not completely self-consistent, also provides a reasonable physical picture for the magnetic moment of the electron.

The analysis above has shown that this rotating phasor is mathematically equivalent to the usual complex Schrödinger equation.  Is it really just a matter of preference which representation we choose?  Not entirely, because a real physical rotation, with a definite frequency and spatial fine structure, should be measurable.  If one probes the behavior of electrons at frequencies $\sim 10^{20}$ Hz $= mc^2/h$, particularly with a circularly polarized probe, one should expect to see a sharp resonance in some sort of spectral response, perhaps associated with spin-flip of the electron in a magnetic field.  For example, one could examine the scattering of polarized x-rays in the 511 keV range.   Furthermore, the fine structure of the spin model identified a periodicity on the scale of $2r_c = 2\hbar/mc$, which would correspond to a momentum transfer $\hbar k = \pi mc \sim 1.5$ MeV/c.  Perhaps some of the relevant measurements have already been carried out – I have not surveyed the literature.  It would be interesting to see whether such results have any bearing on the model described in this paper.

To my mind, this real rotating phasor turns the abstractions of a complex wavefunction and spin into concrete reality. This model is so conceptually simple, that it is surprising that something similar does not appear in the early literature of quantum mechanics.  Ohanian [3] comments that indeed, several early researchers had initially considered rotating solid electrons to explain spin, but had dropped these models because of local speeds greater than c.  The hydrodynamic picture illustrated here may provide a way around these earlier difficulties.

This model certainly cannot account for many of the other problems of quantum mechanics – for example, the measurement problem and quantum coherence – but it may help to remove some of the quantum mystery.

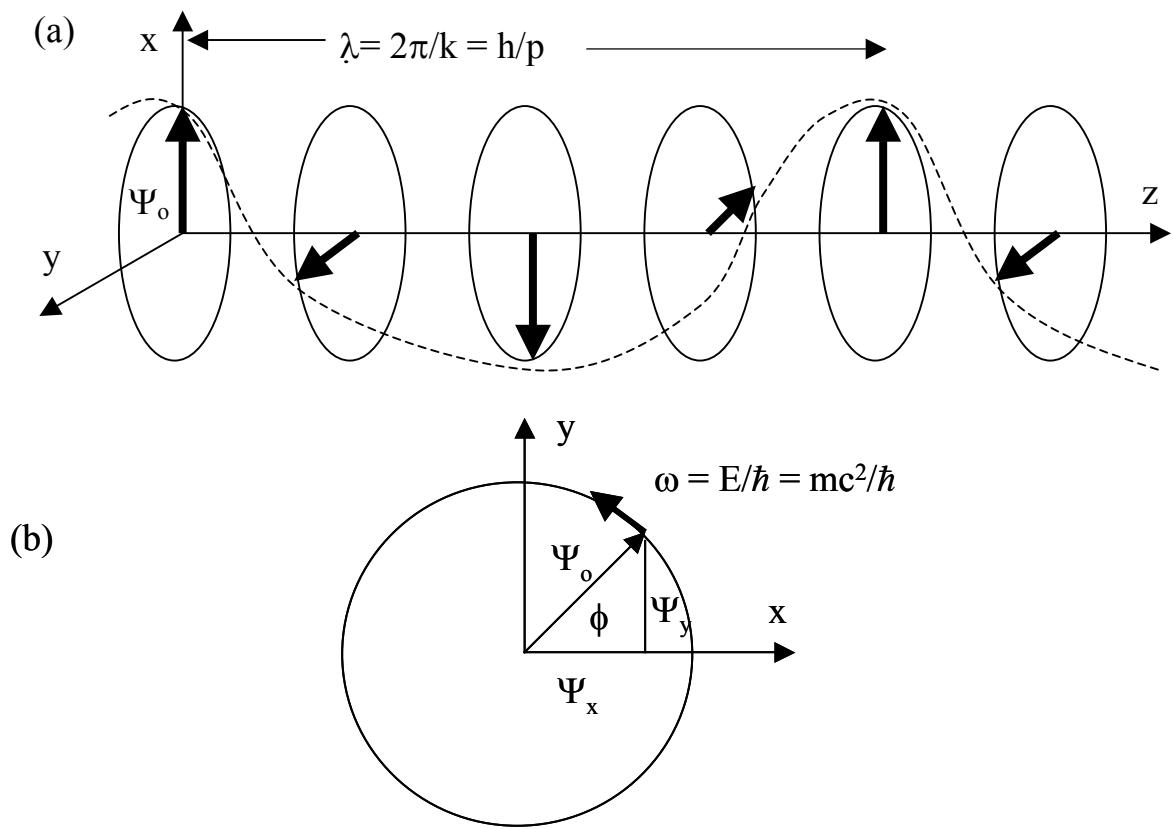

Fig. 1. Picture of real-space helical wave representing electron quantum wavefunction with spin.
(a) Evolution of helix for wave propagating in z-direction.
(b) Rotation of vector phasor for fixed position.

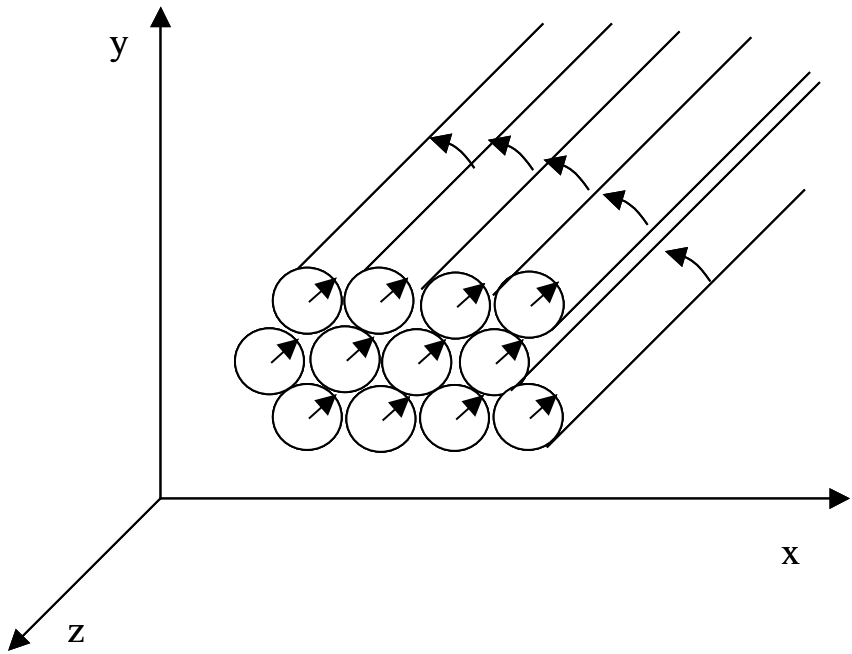

Fig. 2. Picture of hydrodynamic model of electron spin, with parallel array of vortices on scale of $r_c = \hbar/mc = 0.4$ pm.